\documentstyle[12pt]{article}

\large
\newcommand{\nn}{\nonumber}
\newcommand{\be}{\begin{eqnarray}}
\newcommand{\ee}{\end{eqnarray}}

\newcommand{\mat} {\left ( \begin{array}{cc}}
\newcommand{\emat} { \end{array}\right )}
\newcommand{\matt}{\left ( \begin{array}{ccc}}
\newcommand{\ematt}{\end{array} \right )}
\newcommand{\matf}{\left ( \begin{array}{cccc}}
\newcommand{\ematf}{\end{array} \right )}
\newcommand{\vect}{\left ( \begin{array}{c}}
\newcommand{\evect}{\end{array} \right )}
\newcommand{\beq}{\begin{equation}}
\newcommand{\eeq}{\end{equation}}

\newcommand{\bnu} {\overline\nu}
\begin{document}
\setlength{\baselineskip}{17pt}
\pagestyle{empty}
\vfill
\eject
\begin{flushright}
SUNY-NTG-00/69
\end{flushright}

\vskip 2.0cm
\centerline{\Large \bf  
Virasoro Constraints and Flavor-Topology Duality in QCD}

\vskip 1.2cm
\centerline{ D. Dalmazi$^{*}$ and J.J.M. Verbaarschot$^{\dagger}$}
\vskip 0.2cm
\centerline{ $^*$ {\it UNESP , Guaratinguet\'a - S.P., Brazil, 12.500.000}}
\vskip 0.2cm
\centerline{ $^\dagger$ {\it Department of Physics and Astronomy, SUNY, 
Stony Brook, New York 11794}}
\vskip 0.2cm
\centerline{\tt dalmazi@feg.unesp.br ,\, 
verbaarschot@nuclear.physics.sunysb.edu}
\vskip 1.2cm

We derive Virasoro constraints for the zero momentum part of 
the QCD-like partition functions in the sector of topological charge $\nu$. 
The constraints 
depend on the topological charge  only
through the combination $N_f +\beta\nu/2$
where the value of the Dyson index $\beta$ is determined by the
reality type of the fermions. 
This duality between flavor and topology is inherited by the 
small-mass expansion
of the partition function and {\em all}
spectral sum-rules of inverse powers of the eigenvalues of
the Dirac operator. 
For the special case $\beta=2$ but arbitrary topological charge
the Virasoro constraints are solved uniquely by  a Generalized
Kontsevich Model with potential ${\cal V}(X) = 1/X$. 

\noindent
{\it PACS:} 11.30.Rd, 12.39.Fe, 12.38.Lg, 71.30.+h \\ \noindent
{\it Keywords:}Virasoro Constraints; QCD Dirac Spectrum;
Kontsevich; Chiral Random Matrix Theory.
 
\vfill
\noindent

\eject
\pagestyle{plain}

\section{Flavor-Topology Duality}
Through the work of 't Hooft we know that the low-energy limit of QCD
is dominated by  light  flavors
and topology \cite{hooft}. We expect that the same will be the case for the 
low-lying
eigenvalues of the Dirac operator. Indeed, for massless
flavors, the fermion determinant results in the repulsion of eigenvalues
away from $\lambda = 0$. It is perhaps less known that the presence
of exactly zero eigenvalues has the same effect. The reason is
the repulsion of eigenvalues which occurs in all interacting systems
and has probably best been understood in the context of 
Random Matrix Theory where the eigenvalues obey the Wigner repulsion 
law \cite{Wign51}.

In QCD, the fluctuations of the 
low-lying eigenvalues of the Dirac operator are described
by chiral Random Matrix Theory (chRMT) \cite{SV,V,ADMN}. 
This is a Random Matrix Theory
with the global symmetries of the QCD partition function. It is characterized
by the Dyson index \cite{Dyson3}
$\beta$ which is equal to the number of independent 
variables per matrix element. For QCD with fundamental fermions 
we have $\beta =1$ for $N_c=2$ and $\beta =2 $ for $N_c > 2$. For QCD
with adjoint fermions and $N_c \ge 2$ the Dirac  matrix can
be represented in
terms of self-dual quaternions with $\beta = 4$. 
The main ingredient of the chRMT partition function is 
the integration measure which includes the Vandermonde 
determinant. In terms of Dirac eigenvalues $i\lambda_k$
it is given by $\prod_{k<l} |\lambda_k^2 -\lambda_l^2|^\beta 
\lambda_k^{\beta \nu+\beta-1}$. 
Therefore, the presence 
of $ N_f$ massless flavors, with fermion determinant given by 
$\prod_k\lambda_k^{2N_f}$,
has the same effect on eigenvalue correlations
as $\nu
=2N_f/\beta$ zero  eigenvalues. 
More precisely, the joint eigenvalue distribution
only depends on the combination $2N_f + \beta \nu$ \cite{V,Vinst}. 
Based on the conjecture \cite{SV,V}
that the zero-momentum part of the QCD partition function, 
$Z_{N_f,\nu} (m_1, \cdots, m_{N_f})$, is a 
chiral Random Matrix Theory, it was 
suggested \cite{cambridge} that its mass dependence
obeys the duality relation
\be
Z_{N_f,\nu} (m_1, \cdots, m_{N_f})\sim \prod_f m_f^\nu 
Z_{N_f+\nu,0}(m_1, \cdots, m_{N_f}, 0, \cdots, 0).
\label{dual-intro}
\ee
This relation, which is now known as flavor topology-duality,
 is a trivial consequence of the flavor dependence
of the  chRMT joint eigenvalue distribution. 

The mass dependence           
of the chRMT partition function 
can be reduced to a
unitary matrix integral which is known from 
the zero momentum limit of Chiral Perturbation Theory
\cite{GL,LS}. Starting from this representation of the low energy limit
of the QCD partition function, also known as the finite volume partition
function, 
flavor-topology duality 
was first proved  
for $N_f =2$ in \cite{cambridge}. However, its generalization to arbitrary
$N_f$ and $\nu$ 
has only been achieved for $\beta =2$ \cite{AW,ADDV}. The relation 
(\ref{dual-intro}) has been particularly  useful for establishing relations
between  correlation functions of Dirac eigenvalues and finite volume
partition functions 
\cite{ADDV,FV1,FV2,FV3,FV4}.

For $\beta =2$ the unitary matrix integral, 
which represents the low-energy limit
of the QCD partition function, is also known as 
the one-link integral of two-dimensional QCD
(see for example \cite{rossi}) 
or the Brezin-Gross-Witten model \cite{GW}. For zero topological charge
it can also be represented as a generalized Kontsevich model \cite{mironov}
with potential ${\cal V}(X) = 1/X$. 
This model has been  discussed extensively in the context of topological gravity
(for  reviews see \cite{Dijkgraaf,zinn,Dalmazi}.)
In this case the one-link integral  has been analyzed
\cite{mironov,kim} by means of Virasoro constraints which are based on its
invariance properties. 
The main issue we wish to address in this article is whether flavor-topology
duality of the one-link integral can be understood 
without relying on its chRMT representation.
We will do this by deriving Virasoro constraints for arbitrary topological
charge. For $\beta=2$ we find a 
nonperturbative solution of these constraints in the form
of a generalized Kontsevich model. 

In section 2 we will derive the Virasoro constraints for
arbitrary $\nu$ and three different values of $\beta$. 
We observe that they satisfy the flavor-topology duality relations.
A recursive solution
of these relations is presented in section 3.  
 It also provides
us with an efficient derivation of 
sum-rules for the inverse Dirac eigenvalues 
to high order.  In the second part of this section we discuss the uniqueness
of the nonperturbative solution of the Virasoro constraints for $\beta =2$. 
Concluding remarks are made in section 4.

\section{Virasoro Constraints}

In the simplest case, $\beta=2$ and vanishing topological
charge $\nu=0$, the small-mass Virasoro
constraints were first found in \cite{mironov}
after an identification of
the appropriate unitary integral and the corresponding Generalized Kontsevich
Model (GKM).
Here we work entirely in the context of 
unitary integrals and derive a simple form of the 
Virasoro constraints valid for arbitrary topological 
charge and $\beta$. They are obtained by expanding the partition function
in powers of the masses.  

\subsection{{\bf $\beta = 2$}}

The low energy limit of QCD in the phase of broken chiral symmetry is
a gas of weakly interacting Goldstone bosons. Its partition function
is determined uniquely by the invariance  properties of the Goldstone
fields.  
For quark  masses $m$ and space-time volume $V$ in the range 
\be
\frac 1{\Lambda} \ll V^{1/4} \ll \frac 1{m_\pi},
\ee
(where $m_\pi = \sqrt{2m\Sigma/F^2}$ is the mass of the Goldstone modes with 
$F$ the pion decay constant and $\Sigma$ the chiral condensate) 
the partition function for the Goldstone modes 
factorizes \cite{GL,LS} into a zero-momentum part, also
known as the finite volume partition function, and
a nonzero momentum part. 
For  fundamental fermions and a gauge group with $N_c\ge 3$, 
the finite volume partition function in the sector of topological charge
$\nu$ is given by the following integral over the unitary group  
\cite{GL,LS}
\be
{\cal Z}_{\nu}^{\beta=2}({\cal M},{\cal M}^{\dagger})= 
\int_{U\in U(N_f)} 
dU\,  \left(\det U\right)^{\nu} e^{\frac 12 
 {\rm Tr }({\cal M}U^{\dagger} + {\cal M}^{\dagger}U)},
\label{onelink}
\ee
where ${\cal M}=MV\Sigma$. Here and below we always take $\nu \ge 0$.
The quantity $M$ stands for  the original 
unscaled quark mass matrix.
The partition function is normalized such that 
${\cal Z}_{\nu}^{\beta=2}({\cal M}, {\cal M}^\dagger ) \to 
{\det }^\nu({\cal M})$ for ${\cal M} \to 0$. 

Especially for $\nu=0$, the integral (\ref{onelink}) 
has been  studied extensively in the literature.  
In the context of lattice QCD it is 
known as the one-link integral  \cite{GW,Brower,mironov}  
(see also  \cite{rossi} for a review).
Of particular interest is the fact that 
for $N_f\to\infty$ the partition function
${\cal Z}_{\nu}^{\beta=2}$ undergoes a phase 
transition \cite{GW} from a small mass phase ( ${\cal M}\to 0 $)
to a large mass phase ( ${\cal M}\to\infty $ ). 
The large mass expansion is asymptotic and its coefficients 
can be calculated, even for finite $N_f$,  
by expanding about a flavor-symmetric saddle-point. 
For ${\rm Re} ({\cal M})\to\infty $
such expansion can be 
carried out  in a efficient  way by means of the 
the (large-mass) Virasoro constraints found in \cite{GN}.
For ${\rm Im} ({\cal M})\to\infty $, at fixed value of ${\rm Re} ({\cal M})$, 
other flavor-nonsymmetric saddle points contribute to the partition
function as well, 
and it is not known whether the expansion can  be
carried out by means of Virasoro constraints
(for an explicit calculation including nonsymmetric saddle-points
see \cite{DJ}).
On the other hand, the small-mass expansion has a non-zero radius of 
convergence, and it can be determined by different methods for
any finite $N_f$. For example, one can expand the exponential in (\ref{onelink})
and calculate the corresponding unitary integrals 
systematically using 
a character expansion \cite{bars,balantekin}. Another 
efficient technique is again the use of 
the (small-mass) Virasoro constraints found 
in \cite{mironov,kim,ADDV}. 
By generalizing such  constraints
to arbitrary topological charge $\nu$ we will show
that they naturally lead to flavor-topology duality.

First of all, using the unitary invariance of the measure
in (\ref{onelink}) we can deduce 
the covariance properties of the partition function
under a redefinition of the mass matrix
${\cal M}\to V^{-1} {\cal M}$,
\be
(\det {\cal M})^{-\nu}{\cal Z}^{\beta=2}_{\nu}({\cal M},{\cal M}^{\dagger})\, 
=\, (\det V^{-1} {\cal M})^{-\nu}\, 
{\cal Z}^{\beta=2}_{\nu}(V^{-1}{\cal M}, {\cal M}^{\dagger}V).
\label{covariance}
\ee
This  equation implies that 
$(\det {\cal M})^{-\nu}{\cal Z}^{\beta=2}_{\nu}({\cal M},{\cal M}^{\dagger})$
is a  symmetric function of the eigenvalues 
of the hermitean matrix $L_{ab}\equiv ({\cal M}{\cal M}^{\dagger})_{ab} $.
Following \cite{mironov} we can introduce the
infinite set of variables  
\be 
t_k=\frac 1k {\rm Tr}\left(\frac{{\cal M}{\cal M}^{\dagger}}{4}\right)^k
\, , \quad k\ge 1, 
\label{tk2}
\ee
which are explicitly symmetric with respect to  permutations of 
the eigenvalues and write 
\be
(\det {\cal M})^{-\nu}{\cal Z}^{\beta=2}_{\nu}({\cal M},{\cal M}^{\dagger})\, 
= \,  G_{\nu}(t_k)\, ,
\label{Ansatz}
\ee
where $G_{\nu}(t_k)$ has the Taylor expansion
\be 
G_{\nu}(t_k)\, = \, 1 + a_1 t_1 + a_2 t_2 + a_{11} t_1^2 + \cdots \, .
\label{taylor0}
\ee
A simple consistency check on (\ref{Ansatz}) is that 
unitary integrals
are only nonvanishing if the powers of $U$ and $U^{\dagger}$ are
equal.
Following \cite{mironov} we can find  the coefficients 
of the Taylor expansion (\ref{taylor0}) from the differential
equation 
\be
 \frac{\partial^2{\cal Z}^{\beta=2}_{\nu}}
{\partial {\cal M}_{ba} \partial {\cal M}^{\dagger}_{ac}} 
\, = \, \frac{\delta_{cb}}4{\cal Z}^{\beta=2}_{\nu}  \, , 
\label{djj}
\ee
which follows directly from (\ref{onelink}). AS a consequence, 
$G_{\nu}(t_k)$ satisfies the equation
\be 
\left[ \frac{\partial^2}{\partial {\cal M}_{ba} \partial {\cal M}^{\dagger}_{ac}} +
\nu {\cal M}^{-1}_{ab} \frac{\partial}{\partial {\cal M}^{\dagger}_{ac}}\right]
G_{\nu}(t_k) \, = \, \frac{\delta_{cb}}4 G_{\nu}(t_k).
\label{fdif}
\ee
Using the chain rule 
\be 
\frac{\partial}{\partial {\cal M}^{\dagger}_{ac}}\, &=& \, 
\sum_{k\ge 1}\left[\frac{({\cal M}{\cal M}^{\dagger})^{k-1}
{\cal M}}{4^k}\right]_{ca}\partial_k \, , \nn\\
\frac{\partial}{\partial {\cal M}_{ba}}\, &=& \, 
\sum_{k\ge 1}\left[\frac{{\cal M}^{\dagger}
({\cal M}{\cal M}^{\dagger})^{k-1}}{4^k}\right]_{ab}\partial_k \, ,
\label{chainrule}
\ee
with  $\partial_k= \partial/\partial t_k $ we immediately 
obtain from (7)
\be
\sum_{s=1}^{\infty} \Big (\frac{{\cal M}{\cal M}^{\dagger}}{4} \Big 
)^{s-1}_{cb}
\left\lbrack {\cal L}^{\beta=2}_s  - 
\delta_{s,1} \right\rbrack G_{\nu}(t_k)\,  = \, 0,
\label{mmequation}
\ee
where
\be
 {\cal L}_s^{\beta=2}  \, = \, \sum_{k=1}^{s-1} \partial_k\partial_{s-k} 
+ \sum_{k\ge 1}kt_k\partial_{s+k} + (N_f+\nu )\partial_s \, , \qquad s\ge 1\, ,
\label{virasoro}
\ee
and our convention throughout this paper is that terms like 
the first one  in the expression for
${\cal L}_s$ vanish for $s=1$.
The operators ${\cal L}_s$ obey a sub-algebra of the 
Virasoro algebra 
\be
\left[ {\cal L}_r , {\cal L}_s \right] = (r-s)
{\cal L}_{r+s}
\label{valgebra}
\ee
without central charge and $r,s \ge 1$.

At fixed value $N_f$ the $t_k$ are not independent, and the coefficients
of the expansion (\ref{taylor0}) are not determined 
uniquely\footnote{Of course, they are determined uniquely up to the order
that the $t_k$ are independent.}. 
In (\ref{mmequation})
the matrix elements of ${\cal M}{\cal M}^{\dagger}$
and the $t_k$ are not independent so that we cannot conclude 
from (\ref{mmequation}) that
the coefficients of $({\cal M}{\cal M}^{\dagger})^s$ vanish. 
Indeed, the corresponding equations for $N_f =1$ and $\nu =0$ are inconsistent.
However, in order to fix the coefficients in (\ref{taylor0}) uniquely,
we may supplement  (\ref{mmequation}) with  additional equations. We do that
by requiring that all coefficients of $({\cal M}{\cal M}^{\dagger})^s$ vanish,
\be
\left\lbrack {\cal L}_s^{\beta=2}  
- \delta_{s,1} \right\rbrack G_{\nu}(t_k)\,  = \, 0\, , 
\quad s\ge 1.
\label{vira2}
\ee
This procedure is justified provided that
the  all equations are consistent.  
Indeed, for $N_f \to \infty$ the matrix elements
and the $t_k$ are independent so that (\ref{vira2}) must be valid. 
By inserting the Taylor expansion (\ref{taylor0}) into (\ref{vira2}) 
we obtain an  inhomogeneous set linear 
equations  for the coefficients which depend on the parameter $N_f+\nu$.
Inconsistencies can only  arise for isolated values
of $N_f+\nu$ for which the homogeneous part of the equations becomes
linearly dependent. This is indeed what happens  in the example
$N_f =1$ and $\nu =0$. 

Because of the commutation relations (\ref{valgebra}) the constraints for
$s \ge 3$ follow from the constraints for $s=1$ and $s=2$. 
The coefficients of the 
Taylor expansion (\ref{taylor0}) can be found
recursively by solving the constraints for $s=1$ and $s=2$.
This will be carried out in section 3 after deriving the Virasoro 
constraints for $\beta=4$ and $\beta=1$. 

\subsection{{\bf $\beta=4$}}

Similarly,
we now derive the Virasoro constraints 
for adjoint fermions 
($\beta =4$). For fermions in the adjoint 
representation, the zero-momentum Goldstone 
modes belong to the coset space
$SU(N_f)/SO(N_f)$. These Goldstone fields are conveniently parameterized
by $U U^T$ with $U\in SU(N_f)$. 
In the sector
of topological charge $\nu$,  we thus find the finite volume
partition function \cite{LS,SJ},
\be
{\cal Z}^{\beta=4}
_{\overline\nu}({\cal M}_S,{\cal M}^{\dagger}_S)= 
\int_{U\in U(N_f)} 
dU\,  \left(\det U\right)^{2\overline\nu} e^{\frac 12 
 {\rm Tr }({\cal M}_S(UU^t)^\dagger + 
{\cal M}^{\dagger}_SUU^t)} \, ,\label{zbeta4}
\ee
where $\bnu = N_c\nu $ and ${\cal M}_S=M\Sigma V$. 
In this case the mass matrix
${\cal M}_S$ is an arbitrary  symmetric complex matrix.
From ${\cal M}_S \to V^t {\cal M}_S V$ we obtain 
the transformation law
\be
(\det {\cal M}_S)^{-\bar \nu} 
{\cal Z}^{\beta=4}_{\bnu}({\cal M}_S,{\cal M}^{\dagger}_S)\, =\, 
(\det V^t {\cal M}_S V)^{-\bar \nu} 
{\cal Z}^{\beta=4}_{\bnu}(V^{t}{\cal M}_SV,V^{-1}{\cal M}^{\dagger}_S
(V^t)^{-1}) \, .
\label{covariance4}
\ee
This implies
that 
$({\det} {\cal M}_S)^{-\bnu}{\cal Z}_{\bnu}^{\beta=4}({\cal M}_S,
{ \cal M}^\dagger_S)$ is a symmetric function
of the eigenvalues of ${\cal M}_S{ \cal M}^\dagger_S$. We thus have that
\be
(\det {\cal M}_S)^{-\bnu} {\cal Z}^{\beta=4}_{\bnu}({\cal M}_S,
{\cal M}^{\dagger}_S)\,            
= \, G_{\bnu}^{\beta =4}(t_k^S) =  
1 + a_1^S t_1^S + a_2^S t_2^S + a_{11}^S (t_1^S)^2
                       +\cdots \, ,
\label{Ansatz4}
\ee
where in  analogy with (\ref{tk2}) we have defined
\be
t_k^S=\frac 1k {\rm Tr}\left(\frac{{\cal M}_S{\cal M}^{\dagger}_S}{4}\right)^k
\, ,
\label{tk4}
\ee 
and $G_{\bnu}^{\beta=4}(t_k^S)$ will be determined by the  
differential equation 
\be
\frac{\partial^2{\cal Z}^{\beta=4}_{\bnu}}
{\partial {\cal M}_{S\,ba} \partial {\cal M}^{\dagger}_{S\,ac}} 
\, = \, \delta_{bc} {\cal Z}^{\beta=4}_{\bnu}  \, , 
\label{djjs}
\ee
After substituting the expansion (\ref{Ansatz4})
we obtain a differential equation for $G_{\bnu}^{\beta=4}(t_k^S)$ which can 
be written in the form of the Virasoro constraints,
\be
\left\lbrack {\cal L}^{\beta=4}_s  -  
2\delta_{s,1} \right\rbrack G^{\beta=4}_{\bnu}(t^S_k)\,  = \, 0 \, ,
\qquad  s\ge 1\, ,
\label{vira4}
\ee
where 
\be
 {\cal L}^{\beta =4}_s  \, = \, 2\sum_{k=1}^{s-1} \partial_k^S
\partial_{s-k}^S 
+ \sum_{k\ge 1}kt_k^S\partial_{s+k}^S + 
(N_f + 2\bnu + s )\partial_s^S 
\, ,
\label{virasoro4}
\ee
and $\partial_k^S=\partial / \partial t_k^S$. One 
can easily check that ${\cal L}_s^{\beta=4}$ 
and ${\cal L}_s^{\beta=2}$ satisfy the same algebra.

\subsection{{\bf $\beta=1$}}

For QCD with fundamental fermions and $N_c=2$ the Lagrangian can 
be written in terms 
of fermion multiplets of length $2N_f$ containing
$N_f$ quarks and $N_f$ anti-quarks. The chiral 
symmetry is thus enhanced to $SU(2N_f)$. Since the 
fermion condensate is  anti-symmetric in this enlarged flavor space,
the coset space of the Goldstone
modes is given by  $SU(2N_f)/Sp(2N_f)$. This coset manifold is
parameterized by $UIU^t$ where $U\in SU(2N_f)$ and $I$ is 
the $2N_f   \times 2N_f  $ antisymmetric unit matrix
\be
  \label{I}
I = \left (
\begin{array}{cc} 
0 & {\bf 1}\\
-{\bf 1} & 0
\end{array}
\right ) \, .
\ee
The partition function in the 
sector of fixed topological charge $\nu$ is then given by integrating 
over the group manifold $U(2N_f)$ instead of $SU(2N_f)$,
\be
{\cal Z}_{\nu}^{\beta=1}({\tilde {\cal M}_A},
{\tilde {\cal M}_A}^{\dagger})= 
\int_{U\in U(2N_f)} 
dU\,  \left(\det U\right)^{\nu} e^{\frac 14 
 {\rm Tr }(\tilde{\cal M}^\dagger_AUIU^t + {\tilde{\cal M}_A}
(UIU^t)^{\dagger})},
\label{zbeta1}
\ee
where the mass matrix $\tilde{\cal M}_A=MV\Sigma$ is 
an arbitrary anti-symmetric 
complex matrix. In addition to the usual mass-term given by
\be
\mat 0 & {\cal M} \\ - {\cal M} & 0 \emat
\label{standardmass}
\ee 
it contains di-quark source terms in its diagonal blocks. Below, in
the calculation of the mass dependence of the partition function,
the di-quark source terms will be put equal to zero.
The covariance properties of ${\cal Z}_{\nu}^{\beta=1}$ are given by
\be
(\det\tilde{\cal M}_A)^{-\nu/2}
{\cal Z}_{\nu}^{\beta=1}({\tilde {\cal M}_A},{ \tilde{\cal
M}_A}^{\dagger})
\, = \, 
(\det {V^t \tilde{\cal M}_A V})^{-\nu/2}
{\cal Z}_{\nu}^{\beta=1}({V^t \tilde{\cal M}_A V},
{V^{-1} \tilde{\cal M}_A}^{\dagger} V^{t-1})\, . 
\ee
We thus have that
$(\det \tilde{\cal M}_A)^{-\nu/2}
{\cal Z}_{\nu}^{\beta=1}({ \tilde{\cal M}_A},{ \tilde{\cal M}_A}^{\dagger})$ 
is a symmetric function of the eigenvalues of 
$\tilde{\cal M}_A \tilde{\cal M}^\dagger_A$.
This results in the expansion
\be 
\label{Ansatz1}
(\det  \tilde{\cal M}_A)^{-\nu/2}
{\cal Z}_{\nu}^{\beta=1}({ \tilde{\cal M}_A},
{ \tilde{\cal M}}^{\dagger}_A)\, = \, 
G_\nu^{\beta =1}(t^A_k) = 1 + a^A_1 t^A_1 + a^A_2 t^A_2 + a^A_{11} (t^A_1)^2
+\cdots \, ,
\ee
where
\be 
t_k^A\, &=& \, \frac 1{2k}{\rm Tr}\left(
\frac{{\tilde{\cal M}_A}{\tilde{\cal M}_A}^{\dagger}}{4}\right)^k\, .
\label{tk1}
\ee
The factor  $\frac 12$ in the above definition  
of $t_k^A$ takes into account that $\tilde{\cal M}$ is a  $2N_f\times 2N_f$ matrix
for $\beta=1$. 
 Substituting (\ref{Ansatz1}) in the differential equation 
\be
\frac{\partial^2{\cal Z}^{\beta=1}_{\nu}}
{\partial {\tilde {\cal M}}_{A\,ba} 
\partial {\tilde {\cal M}}^{\dagger}_{A\,ac}} 
\, = \, \frac{\delta_{bc}}{4}{\cal Z}^{\beta=1}_{\nu}  \quad , 
\label{djj1}
\ee
we deduce the differential equations for  
$G_{\nu}^{\beta=1}(t_k^A)$ in the form of Virasoro constraints:
\be
({\cal L}^{\beta=1}_s  - \frac{\delta_{s,1}}2)G_{\nu}^{\beta=1}(t_k^A)\, = \, 0,
\label{vira1}
\ee
with
\be
 {\cal L}^{\beta =1}_s  \, = \, \frac 12\sum_{k=1}^{s-1} \partial_k^A
\partial_{s-k}^A 
+ \sum_{k\ge 1}kt_k^A\partial_{s+k}^A + (N_f +\frac{\nu}2 -\frac s2 )
\partial_s^A \,,
\quad 
\label{virasoro1}
\ee
and $\partial_k^A=\partial/\partial t_k^A$. 
The ${\cal L}^{\beta =1}_s$ satisfy the same algebra as the two other
values of $\beta$ discussed in previous sections. 

\section{Solving the Constraints}

In the first part of this section we discuss the recursive solution of
the Virasoro constraints. Flavor-topology duality and nonperturbative
solutions are discussed in the second part of this section.

\subsection{Recursive Solution of Virasoro Constraints}

Remarkably,
in all three cases, $\beta=1$, $\beta =2$ and $\beta=4$, the 
constraints (\ref{vira2}) , (\ref{vira4}) 
and  (\ref{vira1})
can be written in the  unified form:
\be
({\cal L}^{\beta}_s  - \frac 1{\gamma}\delta_{s,1})G(\alpha ,
\gamma)\, = \, 0
\, ,
\label{virasoro124}
\ee
where (see (\ref{virasoro}), (\ref{virasoro4}) and (\ref{virasoro1}))
\be
{\cal L}^{\beta}_s \, =  \frac 1{\gamma}\sum_{k=1}^{s-1} \partial_k
\partial_{s-k}
+ \sum_{k\ge 1}kt_k\partial_{s+k} + 
\frac 1{\gamma} \left[\alpha +\gamma (2-s) +s-1\right]\partial_s \, ,
\label{vira124}
\ee
and, motivated by
ChRMT results of \cite{jselberg}, we have introduced the notation
 \be
\alpha \, &=& \, (N_f -2)2/\beta +\nu +1 \quad , \nn\\
\gamma \, &=& \, 2/\beta .  
\label{ab}
\ee
The $t_k$ are defined as in (\ref{tk2}), (\ref{tk4}) and
(\ref{tk1}) for $\beta=2,\, 4, \, 1$, respectively.
From (\ref{virasoro124}) we conclude                
that in all three cases the Virasoro constraints only depend
on $N_f$ and $\nu$ through the combination 
$N_f+\beta\nu/2 \,\, (\nu\to\bnu\quad {\rm for}
\quad \beta=4)$. 
Proceeding further, we can recursively solve (\ref{vira124})
by substituting the 
Taylor expansion of the
form (\ref{taylor0}) for $G(\alpha,\gamma)$ in 
(\ref{virasoro124}). By treating the $t_k$ as independent variables,
we find the following relations between the expansion coefficients:
\be
\gamma \sum_{k=1}^\infty 
k(n_{k+1}+1) a_{n_1\cdots n_{k}-1 n_{k+1}+1 n_{k+2} \cdots}
+(\alpha +\gamma) (n_1+1) a_{n_1+1 n_2 \cdots} = a_{n_1 n_2 \cdots}
\label{virs1}\, ,
\ee
for $s=1$ and
\be
(n_1+2)(n_1+1)a_{n_1+2n_2 \cdots} &+& \gamma \sum_{k=1}^\infty 
k(n_{k+2}+1) a_{n_1\cdots n_{k}-1 n_{k+1} n_{k+2}+1 \cdots} \nonumber \\ &+& 
(\alpha +1) (n_2+1) a_{n_1 n_2+1 n_3 \cdots} =0\, ,     
\label{virs2}
\ee
for $s=2$. (The subscript $n_1n_2 \cdots$ of the coefficients $a_{n_1n_2\dots}$
is a shorthand for the partition $1^{n_1}2^{n_2} \cdots$.) 
All higher order Virasoro constraints are satisfied trivially 
through the Virasoro algebra.
If we denote the level of the coefficients by $n \equiv \sum_k k n_k$
and and the total number of partitions of $n$ by $p(n)$,
the total number of unknown coefficients at level $n+2$ is equal to 
$p(n+2)$ whereas 
the total number of inhomogeneous equation
(for $s=1$) is equal $p(n+1)$ and
the total
number of homogenous equations (for $s=2$) is equal to $p(n)$.
The total number of  partitions satisfies the recursion relation 
\cite{combina}
\be
p(n+2) = &&p(n+1) + p(n)  
-p(n-3) -p(n-5) + \cdots \nonumber \\ &-&
 (-1)^k p(n+2-\frac{3k^2-k}2) - (-1)^k p(n+2-\frac{3k^2+k}2)+\cdots \, . 
\ee            
Since the number of 
partitions is a monotonic function of $n$, we  have that
\be
p(n+2) \le p(n+1) +p(n),
\ee
and the number of equations is always larger than 
or equal to the number of coefficients. In Table 1 we give the total 
number of coefficients and the total number of equations up to level $n=10$.
\begin{table}
\caption[]{The total number of unknown coefficients ($P(n)$) and the 
total number of equations ($P(n-1) +P(n-2)$) at level $n$.}
\vspace*{0.3cm}
\renewcommand{\arraystretch}{2}
\centering
\begin{tabular}{|c|c|c||c|c|c|}
\hline
$n$ & $P(n)$ & $P(n-1)+P(n-2)$ & $n $& $P(n)$ & $P(n-1)+P(n-2)$ \\
\hline
1  &  1 &  1    &6    & 11  & 12 \\[-0.4cm]
2  &  2 &  2    & 7   & 15  & 18 \\[-0.4cm]
3  &  3 &  3    & 8   & 22  & 26 \\[-0.4cm]
4  &  5 &  5    & 9   & 30  & 37 \\[-0.4cm]
5  &  7 &  8    & 10  & 42  & 52 \\
\hline 
\end{tabular}
\label{tab:pq}
\end{table}

From normalization condition $a_0 =1$ we conclude that
$a_1=1/(\alpha +\gamma)$. At level $n=2$ we obtain one equation each from
(\ref{virs1})  and (\ref{virs2}), respectively,
\be
(\alpha +1)a_2 + 2a_{11} &=& 0 \, , \nn\\
\gamma a_2 + 2(\alpha +\gamma )a_{11} &=& a_1 = 1/(\alpha +\gamma) \, , \ee
which are solved by
\be
a_{11}\, &=& \, \frac{(1+\alpha)}{2\alpha 
(\alpha +\gamma)(\alpha +\gamma +1)} \, , \nn\\
a_{2}\, &=& \, \frac{-1}{\alpha 
(\alpha +\gamma)(\alpha +\gamma +1)}\, .
\ee
Up to order $({\cal M}{\cal M}^{\dagger})^4$ we find
\be 
&&{\cal Z}^{\beta}_{\nu , N_f}({\cal M}, {\cal M}^\dagger) = 
(\det {\cal M})^{\nu}\times \nn\\
&&\Big[ 1+ 
\frac{{\rm Tr}({\cal M}{\cal M}^{\dagger})}{4(\alpha + \gamma )}
+ \frac{(\alpha +1)({\rm Tr}{\cal M}{\cal M}^{\dagger})^2}
{32\alpha (\alpha +\gamma)(\alpha +\gamma+ 1)} 
- \frac{{\rm Tr}({\cal M}{\cal M}^{\dagger})^2}
{32\alpha (\alpha +\gamma)(\alpha +\gamma+ 1) }  \nn\\
&&+\frac{2{\rm Tr}({\cal M}{\cal M}^{\dagger})^3}{4^33\alpha
(\alpha^2-\gamma^2)(\alpha + \gamma +1)(\alpha + \gamma +2)} - 
\frac{(\alpha -\gamma +2)
{\rm Tr}({\cal M}{\cal M}^{\dagger}){\rm Tr}({\cal M}
{\cal M}^{\dagger})^2}{4^32\alpha
(\alpha^2-\gamma^2)(\alpha + \gamma +1)(\alpha + \gamma +2)}\nn\\
&&+\frac{\left[(\alpha - \gamma)(\alpha +3) +2\right]\left[
{\rm Tr}({\cal M}{\cal M}^{\dagger})\right]^3}
{4^3 6\alpha
(\alpha^2-\gamma^2)(\alpha + \gamma +1)(\alpha + \gamma +2)} \nn\\
&&+\frac{(\gamma-5\alpha -6)
{\rm Tr}({\cal M}{\cal M}^{\dagger})^4}{4^5\alpha
(\alpha +1)(\alpha^2-\gamma^2)(\alpha + \gamma +1)
(\alpha + \gamma +2)(\alpha + \gamma +3)(\alpha -2\gamma)}\nn\\ 
&&-\frac{\left[2\alpha^2 -4\alpha(\gamma -2)-7\gamma +6\right]
{\rm Tr}({\cal M}{\cal M}^{\dagger}){\rm Tr}({\cal M}
{\cal M}^{\dagger})^3}{4^43\alpha
(\alpha +1)(\alpha^2-\gamma^2)(\alpha + \gamma +1)
(\alpha + \gamma +2)(\alpha + \gamma +3)(\alpha -2\gamma)}\nn\\
&&+\frac{\left[\alpha^2+\alpha(5-3\gamma)+2\gamma^2-\gamma +6\right]
\left[ {\rm Tr}({\cal M}{\cal M}^{\dagger})^2\right]^2}
{4^52\alpha
(\alpha +1)(\alpha^2-\gamma^2)(\alpha + \gamma +1)
(\alpha + \gamma +2)(\alpha + \gamma +3)(\alpha -2\gamma)}  \nn\\
&&+\frac{\left[ 3\alpha^2(\gamma -2)-\alpha^3-6 -4\gamma^2+13\gamma +
\alpha (-11 +14\gamma -2\gamma^2)\right]
({\rm Tr}{\cal M}{\cal M}^{\dagger})^2{\rm Tr}({\cal M}
{\cal M}^{\dagger})^2}{4^5\alpha
(\alpha +1)(\alpha^2-\gamma^2)(\alpha + \gamma +1)
(\alpha + \gamma +2)(\alpha + \gamma +3)(\alpha -2\gamma)} \nn\\
&&+\!\frac{
\left[6+\alpha^4+\alpha^3(7-3\gamma)-25\gamma +18\gamma^2 +\alpha^2(
17-21\gamma+2\gamma^2)+\alpha (17-43\gamma +14\gamma^2)\right]
({\rm Tr}{\cal M}{\cal M}^{\dagger})^4}{4^56\alpha
(\alpha +1)(\alpha^2-\gamma^2)(\alpha + \gamma +1)
(\alpha + \gamma +2)(\alpha + \gamma +3)(\alpha -2\gamma)}\nn\\ 
&&+\,  
{\cal O}(({\cal M}{\cal M}^{\dagger})^6) \Big ]\, .
\label{taylor124}
\ee
For $\beta=1$ the mass matrix $\tilde{\cal M}$ has been expressed in terms
of the standard $N_f \times N_f$ mass matrix  
that  occurs in the QCD partition function (see eq. (\ref{standardmass})). 
 For $\beta= 2$ and $\beta =4$ 
the matrix ${\cal M}$ is a complex $N_f\times N_f$ matrix 
as well, but for $\beta =4$ it is symmetric. 
We have checked that 
all coefficients in (\ref{taylor124}) agree with 
the corresponding expansion  for $\beta=2$ obtained in \cite{kim}.
When the masses are degenerate $({\cal M}{\cal M}^{\dagger})_{ab}=
\mu^2 \delta_{ab}$
exact expressions for ${\cal Z}_{\nu,N_f}^{\beta}$  
are known for all three values of the Dyson index. For instance, 
in the special case $N_f=2$ 
and $\beta=1$ the finite volume partition function is given by \cite{SJ} 
\be
{\cal Z}_{\nu , 2}^{\beta =1} ~=~ 2  
\sum_{k=0}^\infty \left( \frac {\mu^2}4 \right)^{({\nu}  +k)}
\frac {[2({\nu}  +k)]!}{k! ({\nu}  + k)! ({\nu}  + k +2)!
(2{\nu}  + k)!} \, ,
\label{exactnf2}
\ee
which is a convergent series for
${\cal Z}_{\nu , 2}^{\beta =1}$.
The reader can check that (\ref{taylor124})
only differs from  (\ref{exactnf2})
by the overall normalization factor 
$2^{1-2\nu}/(\nu! (\nu +2)!)$.
We have also made similar checks for $\beta=4$.  The poles that
appear
in the expansion (\ref{taylor124}) at finite $N_f$, the so called
de Wit-'t Hooft poles \cite{hooftwit}, cancel after rewriting the
expansion in powers of products of the eigenvalues of ${\cal
M}^\dagger {\cal M}$. They can be regulated by choosing a noninteger
value for the topological charge \cite{kim,Poulpr}.

The expansion in terms
of traces of the quark mass matrix can be 
directly compared with $\langle \prod_{a=1}^{N_f}  
\det (D +m_a) \rangle_{\rm Yang-Mills}$ from which we
derive spectral sum-rules for inverse powers of 
the eigenvalues of the Dirac operator \cite{LS}. 
With $\zeta_k$ related to the eigenvalues $\pm i\lambda_k$ 
of the Dirac  operator through $\zeta_k=V\Sigma \lambda_k$, 
the first six sum rules for $m_a = 0$ are given by
\be 
\left\langle\sum_n \frac{1}{\zeta_n^2}\right\rangle \, &=& \, 
\frac 1{4(\alpha + \gamma)}\,  ,\nn\\ 
\left\langle\sum_n \frac{1}{\zeta_n^4}\right\rangle \, &=& \,  
\frac{1}{16\alpha (\alpha +\gamma)(\alpha +\gamma +1)}\,  , \nn\\
\left\langle\left(\sum_n \frac{1}{\zeta_n^2}\right)^2\right\rangle 
\, &=& \, \frac{\alpha +1}{16\alpha (\alpha +\gamma)(\alpha +\gamma +1)} \,  ,
\nn\\
\left\langle\sum_n \frac{1}{\zeta_n^6}\right\rangle \, &=& \, 
\frac 1{32\alpha (\alpha^2 -\gamma^2)(\alpha +\gamma +1)(\alpha +\gamma +2)}
\,  ,\nn\\ 
\left\langle\sum_n \frac{1}{\zeta_n^2}\sum_m \frac{1}{\zeta_m^4}
\right\rangle \, &=& \, 
\frac {\alpha -\gamma +2}
{64\alpha (\alpha^2 -\gamma^2)(\alpha +\gamma +1)(\alpha +\gamma +2)}\,  ,\nn\\ 
\left\langle\left(\sum_n \frac{1}{\zeta_n^2}\right)^3
\right\rangle \, &=& \, 
\frac {(\alpha -\gamma )(\alpha +3) +2}
{64\alpha (\alpha^2 -\gamma^2)(\alpha +\gamma +1)(\alpha +\gamma +2)} \,  .
\label{sumrules}
\ee
The sum 
is over the positive eigenvalues only. 
It is clear that all sum-rules will depend
on the topological charge through the combination 
$N_f+\beta\nu/2$. The first formula in (\ref{sumrules}) 
had already been obtained in \cite{SJ}  from 
the low-energy QCD finite volume partition functions.
For $\beta=2$ our sum-rules agree
with the ones derived in \cite{kim} obtained by means of Virasoro
constraints
{\it and} flavor-topology duality.
The remaining sum-rules have never been derived
before from the finite volume partition functions 
for $\beta =1,4$, but  all except the fourth one have been obtained 
from ChRMT by means of Selberg integrals 
for arbitrary $\beta$ \cite{jselberg}.

\subsection{Flavor-Topology Duality}

So far for the sum-rules, we now return to 
the flavor-topology duality relations which is our
main result. To this end
we construct  a $(N_f+\beta\nu/2)\times (N_f+\beta\nu/2)$ matrix
${\overline {\cal M}}$ from the $N_f\times N_f$
original mass-matrix ${\cal M}$ by adding zeros. 
We thus have that ${\rm Tr}({\cal M}{\cal M}^{\dagger})^k=
{\rm Tr}({\overline {\cal M}}\,\,{\overline {\cal M}}^{\dagger})^k$.
The flavor-topology 
duality relation then follows from the observation that
all coefficients
in the small-mass expansion (\ref{taylor124})
depend on the number of flavors 
explicitly through the 
combination $N_f+ \beta\nu/2$ and can be written as
\be
{\cal Z}_{{\nu},N_f}^{\beta}({\cal M},{\cal M}^{\dagger})&=& 
(\det {\cal M} )^{\nu}
{\cal Z}_{0,N_f+\beta\nu/2}^{\beta}({\overline {\cal M}},
{\overline {\cal M}} )\, ,
\label{ftdual}
\ee
where $\nu \to \bnu $ for $\beta=4$.
For the cases where $\beta\nu/2$ is not an integer,
the flavor-topology duality should be understood as an analytical
continuation\footnote{The analytical continuation in $\nu$ is subtle. 
The correct continuation is obtained by \cite{DJ,ADDV} expressing  
the finite volume partition function in terms of modified Bessel functions 
and make an analytical continuation in the index of the Bessel functions.} 
in $\nu$.
Strictly speaking, we have only shown
the duality relation (\ref{ftdual}) for the recursion relations for
the coefficients of the small-mass expansion of the
partition functions. To complete our proof we have to show that the
Virasoro constraints uniquely determine all coefficients and that
the expansion is a convergent series.
The convergence of the
small mass expansion follows immediately from the compactness of the
unitary group. Indeed, for $\beta =2$, the small-mass expansion can be 
resummed to the $\tau$-function of a KP-hierarchy \cite{mironov,FV4}.
This result can also be obtained from an 
exact reconstruction  of ${\cal Z}_{\nu}^{\beta=2}$ \cite{balantekin}
from the character expansion \cite{bars} 
which is an alternative to the small-mass
expansion derived here. 
The uniqueness of the solution of the Virasoro constraints
is a much more difficult question \cite{morozov91}. 
We have checked explicitly to the order 
$({\cal M}{\cal M}^{\dagger})^8$ that this
is indeed the case for all values of the parameters. 

For $\beta =2$, however, the   
uniqueness follows from the explicit solution \cite{JSV} 
of the equation 
\be
\frac{\partial^2}{\partial {\cal M}_{ba} \partial {\cal M}_{ab}^\dagger } 
{\cal Z}_\nu^{\beta                         
=2}({\cal M}, {\cal M}^\dagger) = 
\frac{N_f}4 {\cal Z}_\nu^{\beta =2}({\cal M}, {\cal M}^\dagger)\, ,
\label{virtrace}
\ee
obtained from the trace of (\ref{djj})
or the Schwinger-Dyson equation \cite{Brower}
\be
{\cal M}_{cd}^\dagger {\cal M}_{db}
\frac{\partial^2}{\partial {\cal M}_{ba} \partial {\cal M}_{ac}^\dagger } 
{\cal Z}_\nu^{\beta                           
=2} ({\cal M}, {\cal M}^\dagger)= 
{\cal M}_{ba}^\dagger {\cal M}_{ab}
\frac{N_f}4 {\cal Z}_\nu^{\beta =2}({\cal M}, {\cal M}^\dagger) \, ,
\ee
obtained by contracting (\ref{djj})
with ${\cal M}$ and ${\cal M}^\dagger$. In both cases, the invariance of the
partition function can be used to reduce the 
equations to a separable 
differential equation in the
the eigenvalues of ${\cal M}^\dagger {\cal M}$. Because we are looking
for a solution that is symmetric in the eigenvalues satisfying
the boundary condition $G_\nu(0) = 1$, it is determined
uniquely by the solution of the single particle equation which happens
to be the Bessel equation. 

Let us discuss this  in more detail for (\ref{virtrace}). The first step is to
write the partition function as
\be
{\cal Z}_\nu^{\beta =2}({\cal M}, {\cal M}^\dagger) =
\left ( \frac{\det{ \cal M}}{\det {\cal M}^\dagger} \right )^{\nu/2}
\tilde {\cal Z}_\nu(\Lambda)\, ,
\ee
where $\tilde {\cal Z}_\nu(\Lambda)$ is a function of the eigenvalues
$x_k$ 
of ${\cal M}^\dagger {\cal M}$ only 
($\Lambda ={\rm diag}(x_1, \cdots x_{N_f}$)).
It satisfies the differential equation
\be
\left [\frac{\partial^2}{\partial 
{\cal M}_{ba}\partial  {\cal M}_{ab}^\dagger }    
- \sum_k\frac {\nu^2}{4x_k^2} \right ]\tilde {\cal Z}_\nu(\Lambda) = 
\frac{N_f}4 \tilde{\cal Z}_\nu(\Lambda)\, .
\label{virtracer1}
\ee
This equation can 
be further simplified 
by introducing the reduced partition function
\be
z_\nu(\Lambda) = \Delta(x_k^2) \prod_k \sqrt x_k \tilde{\cal 
Z}_\nu(\Lambda)\, ,
\ee
where the Vandermonde determinant is defined by $\Delta(x_k^2) =
\prod_{k< l} (x_k^2 -x_l^2)$. We thus have that $z_\nu(\Lambda)$
is a completely antisymmetric function of the eigenvalues with boundary
condition such that $\tilde{\cal Z}_\nu(\Lambda)\sim \prod_k x_k^\nu$ 
for $x_k \to 0$. The  reduced partition function satisfies the
separable differential equation
\be
\sum_k \left [ \partial^2_k - \frac {4\nu^2-1}{4 x_k^2} 
\right ] z_\nu(\Lambda) = N_f z_\nu(\Lambda) \, .
\ee
The solution can thus be expressed as a Slater-determinant
\be
z_\nu(\Lambda) = {\det}_{kl} [\phi_k(x_l)],
\label{slater}
\ee
where the $\phi_k$ are
the regular solutions of the single particle equation up to terms
that vanish in the determinant. This requires further discussion. The
solution of the single particle equation
\be
D_B \phi_k(x) \equiv 
\left [ \partial^2_x - \frac {4\nu^2-1}{4x^2} 
\right ]\phi_k(x) = \omega_k^2\phi_k(x)\, ,
\ee
is the modified Bessel function
\be
\phi_k(x) = \sqrt x I_\nu(\omega_k x).
\ee
In order to cancel the Vandermonde determinant for $x_k \to 0$ we 
necessarily have that all $\omega_k^2$ are equal and thus $\omega_k =1$.
Because of the antisymmetry of the columns in the Slater determinant
(\ref{slater}) we can allow for 
single particle solutions  
that satisfy the equations         
\be
D_B \phi_1(x) &=& \phi_1(x), \nonumber \\
D_B \phi_2(x) &=& \phi_2(x) + \mu_{21} \phi_1(x),\nonumber \\
D_B \phi_3(x) &=& \phi_3(x) + \mu_{31} \phi_1(x)+\mu_{32} \phi_2(x)\, ,
{\rm etc.} \, .
\ee
By redefining $\phi_3, \phi_4, \cdots$
we obtain the equations (for $k \ge 2$)
\be
D_B \phi_k(x) &=& \phi_k(x) + \mu_{k\, k-1}\phi_{k-1}(x)\, ,
\ee
where the $\mu_{k\, k-1 }\phi_{k-1}(x)$ cancel in the expression for
the Slater determinant. 
The solution of these equations follows immediately from the recursion 
relation
\be
D_B[ x^{k+\frac 12 } I_{\nu+k}(x) ]= x^{k+\frac 12} I_{\nu+k}(x) + 
2k x^{k-\frac 12}I_{\nu+k-1}(x).
\ee
We thus find that 
\be
\phi_k =    x^{k+\frac 12} I_{\nu+k}(x).
\ee
and $\mu_{k\, k-1} = 2k$. Since the Bessel equation has only one regular 
solution, we conclude that this solution is unique. The Virasoro 
constraints together with the boundary conditions determine the 
partition function uniquely and is given by a 
Slater-determinant of modified Bessel        
functions (\ref{slater}) which agrees with the known result for the one-link
intergral \cite{Brower,mironov,LS,JSV}. 

Actually the solution of the Virasoro constraints, 
although not its uniqueness can be obtained in a simpler way 
by identifying them with constraints which are satisfied by the 
Generalized Kontsevich Model (GKM). By changing variables 
from the matrix elements
of ${\cal M}$ and ${\cal M}^{\dagger}$ to the matrix elements  
of the hermitean matrix $L\equiv ({\cal M}{\cal M}^{\dagger}/4)$ 
we can rewrite
(\ref{fdif}) as 
\be
\left[ \frac{\partial}{\partial L_{be}}L_{de}\frac{\partial}{\partial L_{dc}}
\, +\, \nu \frac{\partial}{\partial L_{bc}} \right] G_{\nu}(L)=\delta_{bc}
G_{\nu}(L)\, .
\label{fdif2}
\ee
This equation admits a solution in the form of a 
GKM with potential ${\cal V}(X) = 1/X $ where $X$ is a $N_f\times N_f$  
normal matrix. 
More explicitly we have\footnote{
The contour of integration is such that
the integral converges for all values of $L$, and that an integral
over a total derivative of $X$ vanishes.}, 
\be
G_{\nu}(L)= \int \frac{dX}{(\det X)^{N_f-\nu}} e^{{\rm Tr}\left[ LX + 
\frac 1X \right]}\, ,
\label{gkm}
\ee
which is a simple modification of the $\nu=0$ case studied in \cite{mironov}.
In order to verify that (\ref{gkm}) is a solution
of (\ref{fdif2}) we have to insert a total derivative 
in the integral over $X$ and multiply it by 
a second order differential operator as follows
\be 
\frac{\partial}{\partial L_{be}}\frac{\partial}{\partial L_{dc}}
\int dX\frac{\partial}{\partial X_{ed}}\left(
\frac 1{(\det X)^{N_f-\nu}} e^{{\rm Tr}\left[ LX + 
\frac 1X \right]}\right) \, = \, 0\, .
\ee

Concluding, for $\beta=2$ the Virasoro constraints naturally lead 
to a simple and rigorous proof of flavor-topology duality (\ref{ftdual})
which is an alternative to the proof presented 
in \cite{AW,ADDV}. 
For $\beta=1$ and $\beta= 4$ a direct solution of the Virasoro constraints
is not known, although we believe that further progress 
can be achieved at least in the
case of equal quark masses where exact formulas 
for the partition function are known \cite{SJ}.
For non-degenerate masses the obstacle is the lack
of a generalization of the Itzykson-Zuber formula  
for integrals over unitary matrices. 
Therefore our proof of flavor-topology duality   
for $\beta=1$ and $\beta =4$ is only valid under the assumption of 
the uniqueness of the solution of the Virasoro constraints.  

\section{Conclusions}

In a parameter range where the main contribution 
to the QCD partition function comes
from the zero-momentum component of the Goldstone modes, 
the partition functions
reduce to zero-dimensional integrals
over the group manifolds $SU(N_f)$ , $SU(N_f)/O(N_f)$ and
$SU(2N_f)/Sp(2N_f)$, for fermions
in the fundamental representation and $N_c\ge 3$ ($\beta=2$), 
$N_c=2$ ($\beta=1$), and adjoint fermions
with $N_c\ge 2$ ($\beta=4$), respectively.
In the sector of fixed topological charge $\nu$ we
have shown that these partition functions satisfy 
differential equations in the form of 
Virasoro constraints which determine
the small-mass expansion for arbitrary
values of the topological charge. 
We have calculated the coefficients of the small
mass expansion recursively up to the order
$m^8$ and have obtained new spectral
sum rules for inverse powers of the eigenvalues
of the Dirac operator for $\beta=1$, $\beta=2$  and $\beta=4$. 
For  all three values of  $\beta$ the constraints 
depend on the topological charge through the 
combination $N_f +\beta\nu/2 $ which demonstrates   
flavor-topology duality for the corresponding
small-mass ($m\to 0$) expansions. 
For the special case 
of $\beta=2$ the uniqueness of the solution of the Virasoro
constraints and the convergence of the small mass expansion implies
flavor-topology duality for the
 full (finite $m$) partition function. 
In this case, we have  found a non-perturbative solution of the 
Virasoro constraints in the form of a Generalized
Kontsevich Model for arbitrary topological 
charge $\nu$ which complements the
$\nu=0$ results of \cite{mironov}. We have
not been able to find an analogous solution 
for $\beta=1$ and $\beta=4$.
Based on numerical evidence we believe
that for these two cases the solution of the
Virasoro constraints is unique as well, but a rigorous proof could not be
given.  The existence of exact expressions for 
partition functions 
with degenerate masses $({\cal M}{\cal M}^{\dagger })_{ab}=\mu^2\delta_{ab}$
indicates that further progress is possible.

It is worth commenting that, even for $\beta =2$,
although order by order the character expansion is  equivalent 
to the perturbative  solution of the small-mass Virasoro constraints,
it does not provide us with a natural proof of flavor-topology duality 
(see \cite{balantekin}).
It should also be pointed out that the GKM are known
to satisfy both small-mass and large-mass
Virasoro constraints. Thus, in view of our results,
we also expect to find large-mass constraints  for $\nu \ne 0$ 
at least for $\beta=2$. Indeed we have explicitly obtained
such constraints entirely within the context of 
unitary integrals and they depend quadratically on the topological
charge $\nu$, contrary to the linear
dependence of the small-mass constraints. The generalization
of  the large-mass constraints
to  $\beta=1$ and $\beta =4$ seems to be within reach.
Due to the absence of an explicit $N_f$ dependence 
of the expansion coefficients they 
are the most relevant ones from the point of view of both
the replica limit \cite{kim,DJ,ADDV} and the classification
of  universality classes
\cite{DVV}. In particular, for applications
in 2D-gravity  they correspond to the continuum
Virasoro constraints which appear after the 
double scaling limit. 
\section{{\bf Acknowledgments}}
\vskip 0.5cm

This work was partially supported by the US DOE grant
DE-FG-88ER40388. 
The work of D.D. is supported by FAPESP (Brazilian Agency). Gernot Akemann, Poul Damgaard,
Kim Splittorff and Dominique Toublan are acknowledged for useful discussions.

\end{document}